\documentclass[11pt]{article}
\usepackage{geometry}
\usepackage{alphabeta}  
\usepackage{blindtext}
\usepackage{multicol}
\usepackage{amssymb}
\usepackage{amsmath}
\usepackage{amsfonts}
\usepackage{graphicx}
\usepackage{xcolor}%
\usepackage{enumitem}
\usepackage{enumerate}

\providecommand{\U}[1]{\protect\rule{0.5in}{0.5in}}
\begin{document}
\bigskip\begin{titlepage}
		\vspace{.3cm} \vspace{1cm}
		\begin{center}
			\baselineskip=16pt \centerline{\bf{\Large{Poincare Invariance in Discrete Gravity }%
			}}
			\vspace{1truecm}
			\centerline{\large\bf Ali H.
				Chamseddine$^{1,2}$\ , Mariam Khaldieh$^{1}$ }\ \
			 \vspace{.5truecm}
		\end{center}

\begin{center}
{$^{1}$ \textit{Ludwig Maxmillian University, \\[0pt] Theresienstr.
37, 80333 Munich, Germany}}\\[0pt] {\small{}{}{}{}{}{}{}{}{}} 
\par\end{center}

\begin{center}
{$^{2}$ {\small{}{}{}{}{}{}{}{}{}{}}\textit{Physics Department,
}\\
 \textit{ American University of Beirut, Lebanon}}\\
 
\par\end{center}
  
		\vspace{2cm}
		\begin{center}
			{\bf Abstract}
		\end{center}
A formulation of discrete gravity was recently proposed based on defining a lattice and a shift operator connecting the cells. Spinors on such a space will have rotational $SO(d)$ invariance which is taken as the fundamental symmetry. Inspired by lattice QCD, discrete analogues of curvature and torsion were defined that go smoothly to the corresponding tensors in the continuous limit. In this paper, we show that the absence of diffeomorphism invariance could be replaced by requiring translational invariance in the tangent space by enlarging the tangent space from S$O(d)$ to the inhomogeneous Lorentz group $ISO(d)$ to include translations. We obtain the $ISO(d)$ symmetry by taking instead the Lie group $SO(d+1)$ and perform on it Inonu-Wigner contraction. We show that, just as for continuous spaces, the zero torsion constraint converts the translational parameter to a diffeomorphism parameter, thus explaining the effectiveness of this formulation.
 \end{titlepage}

\section{Introduction}

In a new formulation of discrete gravity [8], it was shown how Einstein's spacetime gravity can emerge in the classical limit from a more fundamental discrete structure. The idea that spacetime could be discrete at the most fundamental level has long been an attractive concept in gravity. Nevertheless, a \textit{manifest} continuous limit to the classical formulation has only been shown in this formalism, where the geometry of the discrete lattice is not a priori defined, in contrast to other formalisms, and encompasses only a finite number of degrees of freedom.

To paint a broad picture, consider a discrete space consisting of elementary cells of minimal size or volume, such that each cell is numbered by an integer, $n^α$. The minimal cell size is naturally thought to be of a Planckian scale, below which the notion of a manifold loses meaning as a consequence of large metric fluctuations. The size of these cells is then set to be equal to one. Accordingly, the points within each cell are indistinguishable, and quantities defined on each cell are taken at the `center'. The geometry of the cell boundaries are only defined insofar as  the number of neighbors each cell can have, such that each arbitrarily shaped cell is surrounded by 2d neighbors, and from that the dimension(d) of the discrete space is hence defined. Finally, the discrete manifold considered is a d-dimensional space with a Euclidean signature. To recover a differentiable manifold in the continuous limit, the volume of each cell shrinks to zero when the limit of the length scale ${\ell^μ\equiv Δx^μ \to 0}$ is taken, and the manifold coordinates are recovered from the cell enumeration as 
\begin{equation}
    x^α \equiv \ell^α \cdot n^α \equiv (\ell^1 n^1, \ell^2 n^2, ..., \ell^d n^d).
\end{equation}

Zooming in, we will now describe the structure of the theory defined on individual lattice cells. On such small scales, the cells are thought to have no differentiable structure, and thereon locally flat. A finite number of degrees of freedom is then assigned to each cell through a function $f^a(n)$, with respect to each field involved. To move from one cell to an adjacent one, and given that the cells are orientable, a shift operator $E_n$ is constructed such that it acts on the argument of functions defined in the cell in a forward manner as

\begin{equation}
    \mathbf{E}_β (n) f(n^α) \equiv  f(n^1, ...,n^β+1^β,..., n^d),
\end{equation}
shifting the selected coordinate in the function by one unit, and similarly an inverse operator is defined as $E^{-1}_β \equiv E_{-β}$, which shifts the coordinate in the opposite direction. 

A d-dimensional tangent space is then defined for each cell, on which d-tangent operators are constructed from the shift operators 
\begin{equation}
    \bold{e}_α(n) \equiv \frac{1}{2}(\bold{E}_α(n) - \bold{E}^{-1}_α(n)),
\end{equation}
and to connect every cell with its tangent space, a vielbein (soldering form) is thus respectively defined. 

Since locally the cells are flat, the tangent space then exhibits an $SO(d)$ rotational invariance, and spinors that are defined on the lattice will also have a rotational symmetry. Therefore, using the gauge symmetry of the local rotation group in each cell, and using the spin connection group as a basis, curvature and torsion were  defined inspired by their analogues in lattice gauge theory. To this end, curvature is given by the anti-symmetric expression  
\begin{align}\nonumber \label{Curvature}
    \textit{R}_{μν} (n)&= \frac{1}{2\ell^μ\ell^ν} \Big( \Upsilon_μ(n)\Upsilon_ν(n)\Upsilon^{-1}_μ(n)\Upsilon_ν^{-1}(n) - (μ \longleftrightarrow ν) \Big) \\ 
    & =\frac{1}{2\ell^μ\ell^ν}\Big(Ω_μ(n)Ω_ν(n+1_μ)Ω_μ^{-1}(n+1_ν)Ω_ν^{-1}(n) - (μ \longleftrightarrow ν) \Big)
\end{align}
where 
\begin{equation}
    \Upsilon_\mu (n) = \Omega_\mu(n) E_\mu(n)
\end{equation}
and the spin connection group element is given by 
\begin{equation}
    Ω_μ(n)= e^{ω_μ^{ab}(n) J_{ab}} 
\end{equation}
with $J_{ab}$ as the generators of the symmetry group. 

In line with General Relativity, the zero torsion condition was imposed ($T_{μν}^κ=0$), which enforced that the spin connections $ω_μ^{ab}$ be entirely determined by the soldering forms. Curvature and other gauge invariant quantities were then expressed solely in terms of soldering forms on adjacent cells. 

In the continuous limit, the curvature for the spin connection component is recovered in agreement with the standard result in General relativity
\begin{equation}
    R_{\mu\nu}^{ab} =\partial_{\mu} \omega_{\nu}^{ab}-\partial_{\nu}  \omega_{\mu}^{ab}+\omega_{\mu}^{ac} \omega_{\nu}^{cb}- \omega_{\mu}^{ac} \omega_{\nu}^{cb}
\end{equation}

As one can see, the key feature of this formalism was replacing the smooth manifold of spacetime with a lattice structure, all the while trying to preserve the essential features of General Relativity, notably, diffeomorphism invariance, which was done implicitly. In this work, we replace this invariance with translational one, by considering an extension of the formulation, and expanding the symmetry group to include translations.

\section{Diffeomorphism Invariance in General Relativity} 
The approach of recovering the curvature expression for the discrete manifold previously outlined was simply guided by the 'gauging procedure' in relation to the only symmetry group imposed in that case, namely the rotational group. However, it is known that Einstein's theory of gravity could be described as a gauge theory of the Poincare Group, albeit differently from the standard gauge theories of elementary interactions. The reason for that is the Poincare symmetries of spacetime could be regarded as global symmetries only in the absence of gravity. There one can consider the local $SO(d)$ symmetry as the isometry of the local metric, $g_{μν}=η_{μν}$, where the generators for these spacetime rotations (Lorentz transformations) are 
\begin{equation}
J_{ab}=\zeta_{[  ab]  }^{\mu}\partial_{\mu}=(  x_{a}\delta
_{b}^{\mu}-x_{b}\delta_{a}^{\mu})  \partial_{\mu}%
\end{equation}
and by expanding the symmetry group of the tangent space from $SO(d)$ to $ISO(d)$, we recover the translational symmetry part of the Poincare group, which is generated by 
\begin{equation}
    P_a = ζ^μ_a\partial_μ = δ^μ_a \partial_μ. 
\end{equation}

It is clear that the symmetries described this in formulation for a flat spacetime differ from those expected in Einstein's theory when gravity is present. In this case the symmetry of spacetime is described by the general coordinate transformations, or diffeomorphism invariance, and local rotational symmetry on the tangent space, or local Lorentz transformations, as shown earlier. By gauging the Poincare Algebra, and thereby identifying the translational gauge field and rotational gauge field with the soldering forms $e^a_μ$ and spin connections $ω_μ^{ab}$ respectively, the local symmmetries are realized through the following infinitesimal transformations
\begin{align} \delta e_{\mu}{}^a & = \partial_{\mu}\zeta^a + \omega_{\mu}{}^{ab} \zeta^b - \lambda^{ab}e_{\mu}^b \,, \nonumber\\ \delta \omega_{\mu}{}^{ab} & = \partial_{\mu}\lambda^{ab} +  \omega_{\mu}{}^{ac}\lambda^{cb} -\omega_{\mu}{}^{bc}\lambda^{ca}  . \label{The gauge transformations} 
\end{align}
where the spin connections are distinctly transformed by the rotational group with parameter $\lambda_{ab}$.

Under general coordinate transformations $\Tilde{x}^{ν} = x^ν + ζ^ν$, the gauge fields transform as follows 

\begin{equation*} 
\delta' e_{\mu}^a = \zeta^{\nu}\partial_{\nu}e_{\mu}^a +e_{\nu}^a\partial_{\mu}\zeta^{\nu} \label{gcte}
\end{equation*} 
and similarly for the spin connection, where these fields are simply acted upon by the Lie Derivative along the general coordinate transformations vector field parameter $ζ^ν$. 
Finally, the components of the curvatures of local translations and rotations are given by 
\begin{align} 
    T_{\mu\nu}{}^a & = \partial_{\mu}e_{\nu}{}^a – \partial_{\nu}e_{\mu}{}^a +\omega_{\mu}{}^a{}_b e_{\nu}{}^b  -\omega_{\nu}{}^a{}_b e_{\mu}{}^b \,,\nonumber\\ 
    R_{\mu\nu}{}^{ab} & = \partial_{\mu}\omega_{\nu}{}^{ab}-\partial_{\nu}\omega_{\mu}{}^{ab}  + \omega_{\mu}{}^{ac}\omega_{\nu}{}_c{}^b - \omega_{\nu}{}^{ac}\omega_{\mu}{}_c{}^b  \,. \label{PoinCurvatures} 
\end{align}

The subtlety in connecting the Poincare gauge theory with Einsteins gravity lies in replacing the translational symmetry that is absent in GR, by the general coordinate transformations, and this is realized by setting the curvature component related to these translations to zero, i.e. impose a zero torsion constraint. This has two important implications which will be main objectives explored in this paper. The first implication, which was successfully recovered in the original work [8], is that the spin connection field $ω_μ^{ab}$  becomes entirely determined by the soldering forms $e^a_μ$. This constraint is equivalent to demanding that the spin connection is the Levi-Civita connection which is torsion free, as required by General Relativity. Relaxing this constraint, as usually adapted in extensions of GR, allows the spin connection $ω_μ^{ab}$ to be an independent variable in the theory, and leads to appearance of new dynamics involving the coupling of spin matter to gravity. The second implication bridges Poincare gauge theory with gravity by relating the diffeomorphism transformations to  translations. Using the zero torsion condition and defining field dependent parameters $\zeta^{a}(  x) 
=\zeta^{\nu}(  x)  e_{\nu}^{a}(  x) $  and $\lambda'^{ab}(x) = \zeta^{\nu}(x)\omega_{\nu}^{ab}(x)$, one can write
\begin{align*}
\delta e^a_{\mu} &=   \partial_{\mu}(\zeta^{\nu}  e_{\nu}^{a}) + (\zeta^{\nu}  e_{\nu}^{b}) \omega_{\mu}^{ab} -(\zeta^{\nu}\omega_{\nu}^{ab}) e^b_{\mu} \\
&= \delta' e_{\mu}^a + \zeta^{\nu}T_{\mu \nu}^a .
\end{align*}

Thus the diffeomorphism transformation of the vielbein $\delta' e_{\mu}^{\;a}$
is related to the translational and rotational gauge transformations, keeping in mind that torsion should vanish. In other words, one can derive the diffeomorphism transformations of all tensors, using only the transformation of the fields under translations. We perform one more simplification to the inhomogeneous group $ISO(d)$, which we obtain starting from the rotation group $SO(d+1)$ instead of $SO(d)$ and denote the
$SO(d+1)$ generators by $J_{AB}$ such that
\begin{equation}
    [ J_{AB},J_{CD}]  =\delta_{BC}J_{AD}-\delta_{AC}J_{BD}-\delta _{BD}J_{AC}+\delta_{AD}J_{BC}%
\end{equation}
where $A=a,d+1,$ $a=1,\cdots,d$. Denoting $J_{a(  d+1)  }=rP_{a}$,
we get
\begin{align}
    [  J_{a(  d+1)  },J_{b(  d+1)  }]   & =-J_{ab}\\
    [  J_{ab},J_{c(  d+1)  }]   &  =\delta_{bc}J_{a(d+1)  }-\delta_{ac}J_{b(  d+1)  }%
\end{align}
so that in terms of $P_{a}$ we have
\begin{align}
    [  P_{a},P_{b}]   &  =-\frac{1}{r^{2}}J_{ab}\\
    [  J_{ab},P_{c}]   &  =\delta_{bc}P_{a}-\delta_{ac}P_{b}%
\end{align}
and thus recovering the Poincare group by contracting the $SO(d+1)$ group through taking the limit $r\to\infty$. This formulation was applied successfully to the geometric construction of $N=1$ supergravity as a gauge symmetry of the supersymmetry algebra, just after its discovery in $1976$ and still stands as the most elegant derivation of the supergravity action [5].

In the previous work, while the developed discrete theory was explicitly invariant with respect to the local rotation group, diffeomorphism invariance did not hold. It was simply implied by the freedom of enumeration of the lattice cells by a series of integers, which are converted to  coordinates on the manifold, when the continuous limit is taken. In this paper, we will use the strategy outlined in this section in order to replace diffeomorphism invariance by  translational invariance, expanding the tangent space in the recent formulation of discrete gravity, and find modifications of torsion and curvature due to such translational invariance.

\section{Lattice Gravity in extended Tangent Space}

We develop in this section a formulation of discrete gravity in which diffeomorphism invariane is replaced by translational invariance. 

We start by defining the rotation group $SO(d+1)$ on the extended tangent space, the spin connection can then be expanded as

\begin{equation}
    \frac{1}{4}Δx^μ ω_μ^{AB}γ_{AB} = \frac{1}{4}Δx^μ ω_μ^{ab}γ_{ab} + \frac{1}{2r} Δx^μ e^{a}_μ(n)γ_{a}γ 
\end{equation} \normalsize
where $a,b= 1,..., d$, noting that only Latin indices are summed over, and that the following relations hold for the Clifford algebra basis $γ=γ_{d+1}$
\begin{equation}
    γ^2 =1, \ \ \ \{γ,γ^a\} = 0. 
\end{equation}

It follows that
\begin{equation}  \label{Omega}
    Ω_μ(n) = \exp \Big( ω_μ(n) +\frac{1}{r}e_μ(n) \Big)
\end{equation}\normalsize
where we have defined   $ω_μ(n) = \frac{1}{4}Δx^μ ω_μ^{ab}γ_{ab} $ and $e_μ(n)= \frac{1}{2r} Δx^μ e^{a}_μ(n)γ_{a}γ$, \normalsize
and the group element is then given by 
\begin{equation}
    U(n) = exp \Big( λ(n) +\frac{1}{r}ζ(n) \Big)
\end{equation}\normalsize
with  $λ(n) = \frac{1}{4} λ^{ab}(n)γ_{ab}$ and $ζ(n) = \frac{1}{2r} ζ^{a}(n)γ_{a}γ$. \normalsize Under this group action, we define the transformation of the connection  $\Upsilon_μ(n)=Ω_μ(n)E_μ(n)$ \normalsize as
\begin{equation}
    \Upsilon'_μ = U(n)\Upsilon_μ (n) U^{-1}(n),
\end{equation}
and acting with the shift operator we get
\begin{equation} \label{primeOmega}
    Ω'_μ(n) = U(n) Ω_μ(n)  U^{-1}(n +\hat{μ}),
\end{equation}
or written in an expanded form,
\begin{equation} \large
    e^{ω'_μ(n) +\frac{1}{r}e'_μ(n)} = e^{λ(n) +\frac{1}{r}ζ(n)}e^{ω_μ(n) +\frac{1}{r}e_μ(n)}e^{-(λ(n+\hat{μ}) +\frac{1}{r}ζ(n+\hat{μ}))}.
\end{equation}

In order to find the transformations of $ω$ and $e$, obtaining the full expansion of expressions like $e^{ω_μ(n) +\frac{1}{r}e_μ(n)}$ is very complicated, and is given by Zassenhaus formula, which is only known to order 9
\begin{equation}\label{Zassenhaus}
    e^{X+Y} = e^Xe^Y\prod_{n=2}^\infty e^{C_n(X,Y)}
\end{equation}
where $C_n(X,Y)$ is a homogeneous Lie polynomial of degree n in $X$ and $Y$. However, since we are only interested in the result in the limit $r\to \infty$, we can use the result obtianed by Volkin [11], which is valid to all orders, 
\begin{equation}
    e^{\omega(n)+\frac{1}{r}e(n)}=e^{\omega
(  n)  }\Big(   1+\frac{1}{r}(  e-\frac{1}{2!}[\omega,e]  +\frac{1}{3!}[  \omega,[  \omega,e]  ]-\frac{1}{4!}[  \omega,[  \omega,[  \omega,e] ]]  )  +O( \frac{1}{r^{2}}) \Big)
\end{equation} \normalsize

 Although it will be possible to find closed expressions for the above infinite series for the groups $SO(d+1)$, for $d=2,3,4$, the formulas obtained are not transparent. We will thus illustrate the procedure for the case when $d=2$ and  only indicate the steps needed for $d=3,4$ in the appendix. 

\section{Group Compactification}

In this section we find the transformation of the zweibeins in a 2-dimensional space, and then its torsion, starting with the $SO(3)$ group defined on the tangent space. Its Lie group element  $\Omega_{\mu}(n)$ \normalsize is given by the well known formula 
\begin{equation} \label{LieElement}
    {e^{\frac{i}{2}\ell^μ A^i_μ (n) σ_i}}= {\cos \frac{1}{2} \ell^\mu A_\mu(n) + i \frac{A_\mu^i}{A_\mu} \sin \frac{1}{2} \ell^\mu A_\mu(n) \sigma_i}
\end{equation} \normalsize
in which we have
\begin{equation} 
    A_{\mu}^i(n)\sigma_i = \omega_{\mu}(n)\sigma_3 + \frac{2}{r} e_{\mu}^a(n)\sigma_a 
\end{equation} \normalsize
and the group generators given by
$\omega_{\mu}^{ab} \gamma_{ab}=\omega_{\mu}\sigma_3$ and $\omega_{\mu}^{a3} \gamma_{a3}=e_{\mu}^a \sigma_a $. 
Thus, in the limit $r\to\infty$, we can expand the square root of the expression 
\begin{equation}
    (A_{\mu})^2 = (\omega_{\mu}(n))^2 + \frac{4}{r^2} e_{\mu}^a e_{\mu}^a
\end{equation}\normalsize 
as a power series in $r$, where the summation is on the Latin indices only, and find that the magnitude $A_μ$ in (\ref{LieElement}) can be estimated by $ A_{\mu}(n) \simeq \omega_{\mu}(n)$ up to order  $\mathcal{O}(\frac{1}{r^2}) $\normalsize. With that, we finally arrive at the following result for (\ref{Omega}) 
\begin{equation}\label{limitOmega}
    \Omega_{\mu}(n) = e^{\frac{i}{2}\ell^\mu \omega_\mu(n)\sigma_3} + \frac{2i}{r}{\frac{e_{\mu}^a(n)}{\omega_\mu(n)} \sin{\frac{1}{2}\ell^\mu \omega_\mu(n) }\ \sigma_a .} 
\end{equation}\normalsize

Now we can find, in discrete form, the expressions for the transformations of $ω_μ$ and $e^a_μ$ from  $\Omega'_μ(n)= e^{\frac{i}{2}\ell^\mu(\omega'_\mu(n)\sigma_3 + \frac{2}{r}e'^a_{\mu}(n)\sigma_a)}$ \normalsize working the formula given in (\ref{primeOmega}). Using a similar expression to (\ref{limitOmega}) for the group element 
\begin{equation}
    U(n) = e^{\frac{i}{2}λ(n)\sigma_3} + \frac{2i}{r} \frac{ζ^{a}(n)}{λ(n)} \sin{\frac{1}{2}λ(n)} \ \sigma_a 
\end{equation} \normalsize
and expanding only to orders $\frac{1}{r}$, we get using (\ref{primeOmega})
\begin{align} \nonumber
    \Omega'_{\mu}(n) & =  e^{\frac{i}{2}(λ(n)-λ(n+\hat{μ}) +\ell^μ ω_μ(n))\sigma_3} + \frac{2i}{r}\Big( e^{\frac{i}{2}(λ(n)-λ(n+\hat{μ})\sigma_3}\ \Bar{e}^{a}_{\mu}(n) \\ \nonumber
    & + e^{\frac{i}{2}(λ(n+\hat{μ})-\ell^μ ω_μ(n))\sigma_3} 
     \ \Bar{ζ}^{a}(n) +  e^{-\frac{i}{2}(λ(n)+\ell^μ ω_μ(n))\sigma_3} \ \Bar{ζ}^{a}(n+\hat{μ}) \Big) σ_a.
\end{align}
where we defined $\Bar{e}^{a}_{\mu}(n) \equiv \frac{e^{a}_{\mu}(n)}{ω_{\mu}(n)} \sin(\frac{1}{2}\ell^{\mu}\omega_{\mu}(n))$ \normalsize and $\Bar{ζ}(n) \equiv \frac{ζ^{a}(n)}{λ(n)} \sin(\frac{1}{2}λ(n))$ \normalsize to simplify the notation. Equating the above expression with 
\begin{equation}
    \Omega'_μ(n)=  e^{\frac{i}{2}\ell^{\mu}\omega'_{\mu}(n)\sigma_3} 
    + \frac{2i}{r} \frac{e'^{a}_{\mu}(n)}{ω'^{a}_{\mu}(n)} \sin(\frac{1}{2}\ell^{\mu}\omega'_{\mu}(n)) \sigma_a
\end{equation} \normalsize
and comparing $r$ independent parts, we get
\begin{equation}
    \ell^\mu \omega_\mu'(n) = \ell^\mu \omega_\mu(n) + \lambda(n) - \lambda(n + \mu) 
\end{equation}
or equivalently, an explicit discrete transformation for the spin connection 
\begin{equation} \label{primeW}
    \omega_\mu'(n) = \omega_\mu(n) - \Delta_\mu \lambda .
\end{equation}

For the $r$ dependent part we get 
\begin{align} \nonumber \label{primeE1}
    \frac{e_\mu^{\prime a}}{\omega^\prime_\mu(n)}\sin\frac{1}{2}\ell^\mu\omega_\mu^\prime(n) \  \sigma_a & = \Big( e^{-\frac{i}{2}\ell^\mu\Delta_\mu\lambda(n)\sigma_3}\Bar{e}^{a}_{\mu}(n)+ e^{\frac{i}{2}(\lambda(n+\hat{\mu})-\ell^{\mu}\omega_\mu(n))\sigma_3}\Bar{ζ}^{a}(n) \\
    &\hspace{1.2cm}-e^{\frac{i}{2}(\lambda(n)+\ell^\mu\omega_\mu(n))\sigma_3}\Bar{ζ}^{a}(n+\hat{μ}) \Big)\sigma_a
\end{align} \normalsize 
To further simplify the expression for $e_\mu^{\prime a}$, we set the rotation parameter $\lambda \to 0$ as our gauge choice, as we are mostly interested in the transformation of the zweibein under translations, and substitute (\ref{primeW}) into (\ref{primeE1}) to get an exact solution 
\begin{equation}
    e'^a_{\mu}(n)\sigma_a = \Big(e_{\mu}^a(n) + \frac{\omega_\mu(n)}{\sin \frac{1}{2} \ell^\mu \omega_\mu(n)}\big(e^{-\frac{i}{2} \ell^\mu ω_μ(n)\sigma_3 }\zeta_a(n) - e^{\frac{i}{2} \ell^\mu ω_μ(n) \sigma_3 }\zeta_a(n+\mu)\big)\Big)\sigma_a 
\end{equation} \normalsize

We then use the identity  $e^{i\alpha \sigma_3 }\beta_a\sigma_a = (\cos \alpha \beta^a + \sin \alpha \epsilon^{ab} \beta^b)\sigma_a$ \normalsize in the above solution, to express the transformation of the zweibein $e^a_μ$ in terms of real fields as 
\begin{equation} \label{primeE2}
    e'^a_{\mu}(n) = e_{\mu}^a(n) - \ell^\mu \omega_\mu(n) \cot \frac{1}{2} \ell^\mu \omega_\mu(n) \Delta_\mu\zeta(n) - \omega_\mu(n) \epsilon^{ab} \big(\zeta^b(n) + \zeta^b(n + \mu) \big) .
\end{equation} \normalsize
In the above expression we have to substitute the solution of $ω_μ(n)$ as function of $e^a_μ(n)$, however, such substitution could only be done numerically as the torsion equation is a transcendental equation. The best one can do is to develop a perturbative expansion as function of $\ell^μ$. 
The transformation of the metric is then given as follows 
\begin{align} \nonumber
g'_{\mu\nu}(n) = g_{\mu\nu}(n) 
& - \ell^{\mu}\omega_{\mu}(n)\cot\frac{1}{2}\ell^{\mu}\omega_{\mu}(n)\Delta_{\mu}\zeta^{a}(n) e^{a}_{\nu}(n) \nonumber \\ \nonumber
& - \ell^{\nu}\omega_{\nu}(n)\cot\frac{1}{2}\ell^{\nu}\omega_{\nu}(n)\Delta_{\nu}\zeta^{a}(n) e^{a}_{\mu}(n) \nonumber\\
& - \omega_{\mu}(n)\epsilon^{ab}(\zeta^{b}(n) + \zeta^{b}(n+\hat{\mu})) e^{a}_{\nu}(n) \\ \nonumber
& - \omega_{\nu}(n)\epsilon^{ab}(\zeta^{b}(n) + \zeta^{b}(n+\hat{\nu})) e^{a}_{\mu}(n) + O(\zeta^{a})^2 .
\end{align}
\normalsize
Finally, we obtain the expression for the continuous limit of the $e^a_μ$ transformation as $\ell^μ \to 0$
\begin{equation}
    e'^{a}_{\mu} \to e^{a}_{\nu} - \partial_{\mu} ζ^{a} - \omega_{\mu} \epsilon^{ab} ζ^{b}
\end{equation}

Now, we turn to computing the curvature and torsion. Starting with the $SO(3)$ tangent group for a 2-dimensional space, we recall the curvature expression (\ref{Curvature})  below
\begin{equation}
    \Theta_{μν}(n) = \frac{1}{2\ell^μ\ell^ν} \Big(\Omega_μ(n)\Omega_ν(n+\hat{μ})\Omega^{-1}_μ(n+\hat{ν})\Omega_ν^{-1}(n) - μ \longleftrightarrow ν \Big)
\end{equation} \normalsize
and define the element
\begin{equation}
    P_{\mu\nu}(n) \equiv \Omega_{\mu}(n)\Omega_{\nu}(n+\hat{\mu}) \Omega^{-1}_{\mu}(n+\hat{\nu})\Omega^{-1}_{\nu}(n).
\end{equation}
We expand this product up to order $\frac{1}{r}$, where $\Omega_μ(n)$ is given in (\ref{limitOmega}), and after collecting terms, we get 
\begin{equation}
    P_{\mu\nu}(n) = P^{(0)}_{μν} + \frac{2 i}{r} P^{(r)}_{μν} 
\end{equation} \normalsize
with
\begin{equation}
      P^{(0)}_{\mu\nu}(n) =e^{\frac{i}{2}(\ell^\mu\omega_\mu(n) + \ell^\nu\omega_\nu(n+\hat{\mu}) - \ell^\mu\omega_\mu(n+\hat{\nu})-\ell^\nu\omega_\nu(n))\sigma_3}
\end{equation}
and 
\begin{align} \nonumber
    P^{(r)}_{μν} =& \  \big( e^{\frac{i}{2}(-\ell^\nu\omega_\nu(n+\hat{\mu}) + \ell^\mu\omega_\mu(n+\hat{\nu}) + \ell^\nu\omega_\nu(n))\sigma_3}\Bar{
     e}_\mu^a(n) - μ\longleftrightarrow ν \big)\\ 
     & + \big( e^{\frac{i}{2}(\ell^\mu\omega_\mu(n) + \ell^\mu\omega_\mu(n+\hat{\nu}) + \ell^\nu\omega_\nu(n))\sigma_3}\Bar{e}_\nu^a(n+\hat{\mu}) - μ\longleftrightarrow ν \big)
\end{align} 
While the expressions look complicated, they simplify greatly when we compute the curvature  
\begin{equation} 
    R_{\mu\nu}(n)= \frac{1}{\ell^\mu \ell^\nu} (P^{0}_{μν} -P^{0}_{νμ}) = \frac{4i}{\ell^\mu\ell^\nu}\sin{\frac{\ell^\mu\ell^\nu}{2}\big(\Delta_\mu\omega_\nu(n)-\Delta_\nu\omega_\mu(n)\big)}\sigma_3,
\end{equation} \normalsize
and the torsion simply becomes 
\begin{equation}\label{torsion_a}
    T_{μν}^a \sigma_a= \frac{1}{\ell^\mu \ell^\nu} (P^{(r)}_{μν} - P^{(r)}_{νμ} )=  \frac{2}{\ell^\mu \ell^\nu} P^{(r)}_{μν}
\end{equation}
Finally, in the continuous limit $\ell^\mu\to0$, and following steps and relations used for (\ref{primeE2}), we recover the usual equation for curvature 
\begin{equation}
    R_{\mu\nu}\to(\partial_\mu\omega_\nu-\partial_\nu\omega_\mu)
\end{equation}
and torsion 
\begin{equation}
    T_{\mu\nu}^a \to (\partial_\mu e_\nu^a -\partial_\nu e_\mu^a + \epsilon^{ab}\omega_\mu e_{\nu b} -\epsilon^{ab}\omega_\nu e_{\mu b} )
\end{equation}
Let us now compare the torsion obtained by requiring invariance under the Poincare group $ISO(d)$ with the one obtained, as in the Cartan formulation by requiring invariance under the rotation group $SO(d)$ where it is originally defined in [8] by 
 \begin{equation} \label{oldTorison}
    T_{\mu\nu}^{a\,(old)}(n)=\frac{1}{\ell^\mu}(\Upsilon_\mu(n)e_\nu(n)\Upsilon_\mu^{-1}(n)-e_\nu(n))-\frac{1}{\ell^\nu}(\Upsilon_\nu(n)e_\mu(n)\Upsilon_\mu^{-1}(n)-e_\mu(n))
\end{equation} \normalsize
where  $\Upsilon_\mu(n)e_\nu(n)\Upsilon_\mu^{-1}(n)=e^{\frac{i}{2}\ell^\mu\omega_\mu(n)\sigma_3}e_\nu^a(n+\hat{\mu})\sigma_a e^{-\frac{i}{2}\ell^\mu\omega_\mu(n)\sigma_3}$  \normalsize and can be expressed using   (\ref{LieElement}) to rewrite (\ref{oldTorison}) in a useful form 
\begin{equation}
 T_{\mu\nu}^{a\,(old)}(n)=\frac{1}{\ell^\mu}(\cos\ell^\mu\omega_\mu(n)e_\nu^a(n+\hat{\mu}) + \sin\ell^\mu\omega_\mu(n)\epsilon^{ab}e_\nu^b(n+\hat{\mu})-e_\nu^a(n)) - (\mu \leftrightarrow \nu)
\end{equation}
Comparing the above expression with that in (\ref{torsion_a}), which we write below in the useful form
\begin{align}
    T^a_{μν}σ_a =& \frac{2}{\ell^μ\ell^ν} \Big [e^{\frac{i}{2}(-\ell^μ\ell^νΔ_μω_ν(n)+\ell^μ ω_μ(n+\hat{ν}))σ_3 } \Big( \Bar{e}^a_μ(n)\\ \nonumber
    & \hspace{1 cm}- e^{i(-\frac{1}{2}\ell^μ\ell^νΔ_νω_μ(n)+\ell^ν ω_ν(n+\hat{μ}))σ_3 }\Bar{e}^a_μ(n+\hat{ν}) \Big) - (μ\longleftrightarrow ν) \Big ] σ_a,
\end{align}
we deduce that the $ISO(d)$ torsion includes fine corrections to the $SO(d)$ torsion where the zweibein $e_\mu^a(n)$ becomes $\Bar{e}_\mu^a(n)$ with a very small scaling factor $\frac{\sin\frac{1}{2}\ell^\mu\omega_\mu(n)}{\frac{1}{2}\ell^\mu\omega_\mu(n)}<1$. Other corrections depend on $e^{\frac{i}{2}\ell^μ\ell^ν Δ_μ ω_ν(n)σ_3}$ which is very small for fine lattices. The most important corrections come from the factor $e^{\frac{i}{2}(\ell^μ ω_μ(n+ν)-\ell^ν ω_ν(n+μ))σ_3}$ appearing in the zero torsion condition. 

Lattice gravity based on the $SO(d)$ proved itself by giving excellent agreement when applied to well known geometries, like those of spheres and black holes. We would expect that the new defintion of torsion, although more complicated than the simple torsion based on $SO(d)$, should give an even better precision. Although this is a conjecture, a supporting example is that of supergravity, where the vanishing of torsion associated with the supersymmetry algebra (and this includes the translation generator) produces the correct theory of supergravity valid to all orders. Numerical studies using the new form of torsion is needed to give a definitive answer.

The case for higher dimensions turns out to be more involved and will be left for future explorations. The basic idea is that the approach will slightly differ than the one used in the case of $ISO(2)$, and will involve certain integration of the Zassenhaus formula \eqref{Zassenhaus} to find the curvature elements. For an analogous approach to the one applied in this section, we provide more details regarding the basis used for $ISO(3) $ and $ISO(4)$ and an outline for the steps  in the Appendix.

\section{Conclusion}

In the formulation of discrete geometry defined as a cluster of oriented cells at extremely small scales, possibly Planck scale, constructed in such a way that every cell is surrounded by 2d cells, with d being the dimension of the continuous limit of the space. All the cells could be oriented and a shift operator $E_μ$ is defined to the right and left of every cell. Our basic observation is that at such small scales, locally, every cell can be considered to be flat, and the tangent space will exhibit $SO(d)$ rotational invariance. Just as for lattice gauge theories of the strong interaction where the spinors on such a space is taken to have an $SU(3)$ symmetry, the spinors will now have $SO(d)$ rotational symmetry, allowing for the definition of curvature and torsion for these spaces. A vielbein, or soldering form, is defined for every cell, connecting every cell to its tangent space. The advantage of this construction is that it has a manifest continuous limit. Obviously, diffeomorphism invariance does not hold in discrete spaces, although this construction has proved to be extremely accurate in numerical calculations. It is nonetheless desirable to have the values of the curvature scalar of every cell to be independent of the renumeration. In this paper we proposed to extend the rotational symmetry of the tangent space to include translations and thus to take the group to be the inhomogeneous rotation group $ISO(d)$ : In the continuous limit the group $ISO(d)$ can be obtained by contracting the group $SO(d+1)$ taking the radius of the $(d+1)$th direction to infinity. Although this procedure is simple for Lie algebras, it is highly non-trivial for Lie groups because one has to apply the Baker-Campbell-Hausdorf formula. In this paper we have investigated the modification of curvature and torsion due to translational invariance. The main advantage of our formulation, is the manifest continuous limit, in contrast to other formulations such as Regge [17], [19]. The idea of using the idea of the tangent group to relate geometric theories of gravity to gauge theory proved to be very fruitful [5], [6]. Our conclusion is that, in general, there are very small corrections to torsion, and that to insure invariance of the action, there is a need to modify the definition of the curvature tensor in such a way that it will satisfy some form of Bianchi identity. We have derived the form of the transformations of the vielbein and spin-connections as dictated by $ISO(d)$ invariance for $ d = 2$: The resulting modifications to torsion and curvature are of higher order in function of the lattice size, explaining the reason why the numerical calculations are extremely close to those of the continuous limit [12], [13], [14]. The case when $d =3,4$ is rather involved and will be presented elsewhere.

\newpage
\section*{Appendix A: Tangent group $ISO(3)$}
Consider the group $ISO (3)$ obtained from the contraction of $SO (4)$. Take the following gamma matrix representation
\begin{equation}
    \gamma^a = \sigma^a\otimes\tau_2,\hspace{1cm}a=1,2,3,\hspace{1cm}\gamma_5=1\otimes\tau_3
\end{equation}
\begin{equation}  \gamma^{ab}=i\epsilon^{abc}\sigma^c\otimes1,\hspace{1cm}\gamma_5=\gamma_1\gamma_2\gamma_3\gamma_4
\end{equation}
\begin{equation}
    \gamma^4=1\otimes\tau_1,\hspace{1cm}\gamma^a\gamma^4=-i\sigma^a\otimes\tau_3
\end{equation}
where $\sigma^a$ and $\tau^a$ are two sets of Pauli matrices. The spin connection can then be split as follows
\begin{align}\nonumber
    \frac{1}{4}\omega_\mu^{AB}\gamma_{AB}&=\frac{i}{4}\epsilon^{abc}\omega_\mu^{ab}\sigma^c\otimes1 + \frac{1}{2}\omega_\mu^{a4}(-i\sigma^a\otimes\tau_3)\\\nonumber
    &\equiv\frac{i}{2}\omega_\mu^a\sigma^a\otimes1-\frac{i}{2r}e_\mu^a\sigma^a\otimes\tau_3\\
    &=\frac{i}{2}\omega_\mu^{-a}\sigma^a\otimes\frac{1}{2}(1+\tau_3)+\frac{i}{2}\omega_\mu^{+a}\sigma^a\otimes\frac{1}{2}(1-\tau_3)
\end{align}
where
\begin{equation}
    \omega_\mu^{-a}=\omega_\mu^a - \frac{1}{r}e_\mu^a,\hspace{1cm}\omega_\mu^{+a}=\omega_\mu^a+\frac{1}{r}e_\mu^a
\end{equation}
or equivalently
\begin{equation}
    \omega_\mu^a=\frac{1}{2}(\omega_\mu^{+a}+\omega_\mu^{-a})
\end{equation}
\begin{equation}
    \frac{1}{r}e_\mu^a=\frac{1}{2}(\omega_\mu^{+a}-\omega_\mu^{-a})
\end{equation}
and the group $SO(4)$ splits into two independent $SO(3)$ blocks as $\frac{1}{2}(1+\tau_3)$ and $\frac{1}{2}(1-\tau_3)$ are orthogonal. The group $ISO(3)$ is obtained by letting $r\rightarrow\infty$. Note that
\begin{equation}
[\gamma_{ab},\gamma_{c4}]=\delta_{bc}\gamma_{a4}-\delta_{ac}\gamma_{b4}
\end{equation}
\begin{equation}
    \frac{1}{r^2}e_\mu^a e_\nu^b[\gamma_{a4},\gamma_{b4}]=-\frac{1}{r^2}e_\mu^a e_\nu^b\gamma_{ab}\rightarrow0
\end{equation}
We then denote the product
\begin{align}\nonumber
\Omega_μ(n)\Omega_ν(n+\Hat{μ}) =&\Big(\cos\frac{1}{2}\Delta x^\mu\omega_\mu^{\pm}(n) + i\sigma^a\hat{\omega}_\mu^{\pm a}(n)\sin\frac{1}{2}\Delta x^\mu\omega_\mu^{\pm}(n)\Big)\\ 
 &\Big(\cos\frac{1}{2}\Delta x^\nu\omega_\nu^{\pm}(n+\hat{\mu}) + i\sigma^b\hat{\omega}_\nu^{\pm b}(n+\hat{\mu})\sin\frac{1}{2}\Delta x^\nu\omega_\nu^{\pm}(n)\Big)\\
=&A_{\nu\mu}^\pm(n)+iB_{\mu\nu}^{\pm a}(n)\sigma_a
\end{align} 
where
\begin{align}
      A_{\mu\nu}^\pm(n) =& \cos\frac{1}{2}\Delta x^\mu\omega_\mu^{\pm}(n+\hat{\nu})  \cos\frac{1}{2}\Delta x^\nu\omega_\nu^{\pm}(n) \\
      & - \hat{\omega}_\mu^{\pm b}(n+\hat{\nu})\hat{\omega}_\nu^{\pm b}(n)\sin\frac{1}{2}\Delta x^\mu\omega_\mu^\pm (n+\hat{\nu})\sin\frac{1}{2}\Delta x^\nu\omega_\nu^\pm (n) 
\end{align}
and
\begin{align}\nonumber
    B_{\mu\nu}^{\pm a}(n) = & \hat{\omega}_\mu^{\pm a}(n)\sin\frac{1}{2}\Delta x^\mu \omega_\mu^\pm (n)\cos\frac{1}{2}\Delta x^\nu \omega_\nu^\pm (n+\hat{\mu}) \\
    &+ \hat{\omega}_\nu^{\pm a}(n+\hat{\mu})\sin\frac{1}{2}\Delta x^\nu \omega_\nu^\pm (n+\hat{\mu})\cos\frac{1}{2}\Delta x^\mu\omega_\mu^\pm (n)\\ 
    &-\epsilon^{abc}\hat{\omega}_\mu^{\pm b}(n)\sin\frac{1}{2}\Delta x^\mu \omega_\mu^\pm(n)\hat{\omega}_\nu^{\pm c}(n+\hat{\mu})\sin\frac{1}{2}\Delta x^\nu\omega_\nu^\pm(n+\hat{\mu})
\end{align}
We then have 
\begin{equation}
    P_{\mu\nu}^{\pm}(n) = P_{\mu\nu}^{\pm 0}(n)+iP_{\mu\nu}^{\pm a}\sigma_a
\end{equation}
where
\begin{equation}
    P_{\mu\nu}^{\pm 0}(n) = A_{\mu\nu}^\pm(n)A_{\nu\mu}^\pm(n) + B_{\mu\nu}^{\pm a}(n)B_{\nu\mu}^{\pm a}(n) = P_{\nu\mu}^{\pm 0}(n)
\end{equation}
\begin{equation}
      P_{\mu\nu}^{\pm a}(n) = A_{\mu\nu}^\pm(n)B_{\mu\nu}^{\pm a}(n) - A_{\nu\mu}^\pm(n)B_{\nu\mu}^{\pm a}(n) + \epsilon^{abc}  B_{\mu\nu}^{\pm b}(n)B_{\nu\mu}^{\pm c}(n) = -P_{\nu\mu}^{\pm a}(n)
\end{equation}
The three dimensional curvature is given by
\begin{align}\nonumber
    \Theta_{\mu\nu}(n)&= \frac{1}{4}R_{\mu\nu}^{\phantom{\mu\nu}ab}
\gamma_{ab} + \frac{1}{2}R_{\mu\nu}^{\phantom{\mu\nu}a4}
\gamma_{a4}\\\nonumber
    &=\frac{i}{2}R_{\mu\nu}^{\phantom{\mu\nu}a}
\sigma^a\otimes1 - \frac{i}{2r}T_{\mu\nu}^a\sigma^a\otimes \tau_3\\
    &=\frac{i}{2}R_{\mu\nu}^{\phantom{\mu\nu}-a}
\sigma^a\otimes\frac{1}{2}(1+\tau_3)+\frac{i}{2}R_{\mu\nu}^{\phantom{\mu\nu}+a}
\sigma^a\otimes\frac{1}{2}(1-\tau_3)
\end{align}
where
\begin{equation}
    R_{\mu\nu}^{\pm a}= \frac{1}{\Delta x^\mu\Delta x^\nu}P_{\mu\nu}^{\pm a}
\end{equation}
\begin{equation}
R_{\mu\nu}^{\phantom{\mu\nu}a} = \frac{1}{2}\left(R_{\mu\nu}^{\phantom{\mu\nu}+a} + R_{\mu\nu}^{\phantom{\mu\nu}-a}\right),\hspace{1cm} 
T_{\mu\nu}^a = -\lim_{r\rightarrow\infty}r\left(R_{\mu\nu}^{\phantom{\mu\nu}+a} - R_{\mu\nu}^{\phantom{\mu\nu}-a}\right)
\end{equation}

\section*{Appendix B: Tangent group $ISO(4)$}
We have to find the torsion for $ISO (4)$ and to do this, we need the expression
for the operator
\begin{equation}
    P_{\mu\nu}(n) = S_\mu(n)S_\nu(n+\hat{\mu})S_{\mu}^{-1}(n+\hat{\nu})S_\nu^{-1}(n)
\end{equation}
for the group $SO (4)$. We first note that the group $SO (4)$ is isomorphic to
$SU (2) \times SU (2)$ and we can obtain a closed expression of the curvature for each
$SO (3)$ by using the fact that the connection $\omega_\mu$ split into self-dual and anti
self-dual parts
\begin{align}\nonumber
    \omega_\mu &=\frac{1}{4}\omega_\mu^{\phantom{\mu}{AB}}\gamma_{AB}\hspace{.5cm}A,B=1,\cdots,4\\\nonumber
    &=\frac{1}{4}\omega_\mu^{\phantom{\mu}{AB}}\gamma_{AB}\left(\frac{1}{2}(1-\gamma_5)+\frac{1}{2}(1+\gamma_5)\right)\\
&\equiv\frac{1}{4}\left(\omega_\mu^{+AB} + \omega_\mu^{-AB}\right)\gamma_{AB}
\end{align}
As we are considering the Euclidean $SO (4)$ gauge theory, where $A = a,4$ and
$\frac{1}{2}\epsilon^{ABCD}\omega_\mu^{\pm AB}=\pm\omega_\mu^{\pm CD}$ will each have three independent components (for this we use
\begin{equation}
    \gamma_5\gamma_{AB}=-\frac{1}{2}\epsilon_{ABCD}\gamma_{CD}
\end{equation}
with $\gamma_5=\gamma_1\gamma_2\gamma_3\gamma_4$ with the convention $\epsilon_{1234}=1$. With these conventions we
will identify $\omega_{\mu}^{\phantom{\mu}ab}$ with the three dimensional spin connection. Working in the Weyl representation of the gamma matrices, we take
\begin{align}\nonumber
&\gamma^4 = \begin{pmatrix}
 0 & 1\\
 1 & 0
\end{pmatrix},\hspace{.5cm}
\gamma^a = \begin{pmatrix}
 0 & -i\sigma^a\\ i\sigma^a & 0
\end{pmatrix},\hspace{.5cm}
a=1,2,3\\
&\gamma^5 = \begin{pmatrix}
 1 & 0\\
 0 & -1
\end{pmatrix},\hspace{.5cm}
\gamma^{a4} = \begin{pmatrix}
  -i\sigma^a & 0 \\ 0 & i\sigma^a \end{pmatrix},\hspace{.5cm}
\gamma^{ab}=i\epsilon^{abc}\sigma^c\otimes 1_2
\end{align}
Thus we have
\begin{equation}
    \frac{1}{4}\omega_\mu^{\phantom{\mu}AB}(n)\gamma_{AB} = \begin{pmatrix}
 \frac{i}{2}\omega_\mu^{-a}\sigma^a & 0\\
 0 & \frac{i}{2}\omega_\mu^{+a}\sigma^a
\end{pmatrix}\omega_\mu^{\pm a}\equiv\frac{1}{2}\epsilon^{abc}\omega_\mu^{bc}\pm\omega_\mu^{a4}
\end{equation}
and the group elements are
\begin{equation}
    S_\mu(n)=\begin{pmatrix}
 s_\mu^-(n)& 0\\
 0 &s_\mu^+(n)
\end{pmatrix},\hspace{1cm}s_{\mu}^{\pm}(n)=\exp{\left(\frac{i}{2}\ell^\mu\omega_\mu^{\pm a}(n)\sigma^a\right)}
\end{equation}
Next we compute
\begin{equation}
    P_{\mu\nu}(n) = \begin{pmatrix}
 P_{\mu\nu}^{-0}(n)+i P_{\mu\nu}^{-a}\sigma^a& 0\\
 0 & P_{\mu\nu}^{+0}(n)+i P_{\mu\nu}^{+a}\sigma^a
\end{pmatrix}
\end{equation}
where
\begin{equation}
    P_{\mu\nu}^{\pm 0}(n) = A_{\mu\nu}^\pm(n)A_{\nu\mu}^\pm(n) + B_{\mu\nu}^{\pm a}(n)B_{\nu\mu}^{\pm a}(n) = P_{\nu\mu}^{\pm 0}(n)
\end{equation}
\begin{equation}
      P_{\mu\nu}^{\pm a}(n) = A_{\mu\nu}^\pm(n)B_{\mu\nu}^{\pm a}(n) - A_{\nu\mu}^\pm(n)B_{\nu\mu}^{\pm a}(n) + \epsilon^{abc}  B_{\mu\nu}^{\pm b}(n)B_{\nu\mu}^{\pm c}(n) = -P_{\nu\mu}^{\pm a}(n)
\end{equation}
We can write this in the form
\begin{equation}
    P_{\mu\nu}(n)= P_{\mu\nu}^1(n)+ P_{\mu\nu}^5(n)\gamma_5 + \frac{1}{4} P_{\mu\nu}^{AB}\gamma_{AB}
\end{equation}
where
\begin{equation}
   P_{\mu\nu}^1(n) =\frac{1}{2}\left(P_{\mu\nu}^{+0}(n)+P_{\mu\nu}^{-0}(n)\right),\hspace{.5cm} P_{\mu\nu}^5(n) =\frac{1}{2}\left(P_{\mu\nu}^{-0}(n)-P_{\mu\nu}^{+0}(n)\right)
\end{equation}
We can reconstruct $P_{\mu\nu}^{\phantom{\mu\nu}AB}(n), A,B=1,\cdots,4$ from $P_{\mu\nu}^{\pm a}(n)$
\begin{align}  
    P_{\mu\nu}^{\phantom{\mu\nu}a4}(n)=\frac{1}{2}\left(P_{\mu\nu}^{+a}(n)-P_{\mu\nu}^{-a}(n)\right),\hspace{.5cm} P_{\mu\nu}^{\phantom{\mu\nu}ab}(n) =\frac{1}{2}\epsilon^{abc}\left(P_{\mu\nu}^{+c}(n)+P_{\mu\nu}^{-c}(n)\right)
\end{align}
where
\begin{align} 
  R_{\mu\nu}^{\pm a}(n) &= \frac{P_{\mu\nu}^{+a}(n)}{\Delta x^\mu \Delta x^\nu}\\\nonumber
  &=  \frac{2}{\Delta x^\mu \Delta x^\nu}  \left(A_{\mu\nu}^\pm(n)B_{\mu\nu}^{\pm a}(n) - A_{\nu\mu}^\pm(n)B_{\nu\mu}^{\pm a}(n) + \epsilon^{abc}  B_{\mu\nu}^{\pm b}(n)B_{\nu\mu}^{\pm c}(n)\right)
\end{align}
and
\begin{equation}
    \frac{1}{4}R_{\mu\nu}^{\phantom{\mu\nu}AB}(n)\gamma_{AB} = \begin{pmatrix}
        \frac{i}{2}R_{\mu\nu}^{-a}\sigma^a & 0 \\
        0& \frac{i}{2}R_{\mu\nu}^{+a}\sigma^a
    \end{pmatrix}
\end{equation}
where
\begin{equation}
    R_{\mu\nu}^{-a}(n) = \frac{1}{2}\epsilon^{abc}R_{\mu\nu}^{\phantom{\mu\nu}bc}(n) - R_{\mu\nu}^{\phantom{\mu\nu}a4}(n),\hspace{.5cm}
     R_{\mu\nu}^{+a}(n) = \frac{1}{2}\epsilon^{abc}R_{\mu\nu}^{\phantom{\mu\nu}bc}(n) + R_{\mu\nu}^{\phantom{\mu\nu}a4}(n)
\end{equation}

\section*{Acknowledgments}
The work of A. H. C is supported in part by the National Science Foundation Grant No. Phys-2207663. The work of M.K. is supported by the Deutsche Forschungsgemeinschaft (DFG, German Research Foundation) under Germany's Excellence Strategy --EXC--2111--390814868.

\end{document}